\documentclass[runningheads]{llncs}
\usepackage{graphicx}

\usepackage{cite}
\usepackage{amsmath,amssymb,amsfonts}
\usepackage{algorithmic}
\usepackage{graphicx}
\usepackage{textcomp}
\usepackage{xcolor}
\usepackage{listings}
\usepackage{url}
\usepackage{color}
\usepackage[flushleft]{threeparttable}
\usepackage[inline]{enumitem}
\usepackage{multirow}
\usepackage[font=footnotesize]{subfig}
\usepackage{diagbox}
\usepackage{booktabs}

\definecolor{lightgray}{rgb}{0.95, 0.95, 0.95}
\definecolor{darkgray}{rgb}{0.4, 0.4, 0.4}
\definecolor{editorGray}{rgb}{0.95, 0.95, 0.95}
\definecolor{editorOcher}{rgb}{0.5, 0.26, 0} 
\definecolor{editorGreen}{rgb}{0, 0.5, 0} 
\definecolor{orange}{rgb}{1,0.45,0.13}		
\definecolor{olive}{rgb}{0.17,0.59,0.20}
\definecolor{brown}{rgb}{0.69,0.31,0.31}
\definecolor{purple}{rgb}{0.38,0.18,0.81}
\definecolor{lightblue}{rgb}{0.1,0.57,0.7}
\definecolor{lightred}{rgb}{1,0.4,0.5}
\usepackage{upquote}
\usepackage{listings}
\lstdefinelanguage{CSS}{
  keywords={color,background-image:,margin,padding,font,weight,display,position,top,left,right,bottom,list,style,border,size,white,space,min,width, transition:, transform:, transition-property, transition-duration, transition-timing-function},	
  sensitive=true,
  morecomment=[l]{//},
  morecomment=[s]{/*}{*/},
  morestring=[b]',
  morestring=[b]",
  alsoletter={:},
  alsodigit={-}
}

\lstdefinelanguage{JavaScript}{
  morekeywords={typeof, new, true, false, catch, function, return, null, catch, switch, var, if, in, while, do, else, case, break},
  morecomment=[s]{/*}{*/},
  morecomment=[l]//,
  morestring=[b]",
  morestring=[b]'
}

\lstdefinelanguage{HTML5}{
  language=html,
  sensitive=true,	
  alsoletter={<>=-},	
  morecomment=[s]{<!-}{-->},
  tag=[s],
  otherkeywords={
  >,
	<!DOCTYPE,
  </html, <html, <head, <title, </title, <style, </style, <link, </head, <meta, />,
	</body, <body,
	</div, <div, </div>, 
	</p, <p, </p>,
	</script, <script,
  <canvas, /canvas>, <svg, <rect, <animateTransform, </rect>, </svg>, <video, <source, <iframe, </iframe, </video>, <image, </image>, <header, </header, <article, </article
  },
  ndkeywords={
  =,
  charset=, src=, id=, width=, height=, style=, type=, rel=, href=,
  fill=, attributeName=, begin=, dur=, from=, to=, poster=, controls=, x=, y=, repeatCount=, xlink:href=,
  margin:, padding:, background-image:, border:, top:, left:, position:, width:, height:, margin-top:, margin-bottom:, font-size:, line-height:,
  transform:, -moz-transform:, -webkit-transform:,
  animation:, -webkit-animation:,
  transition:,  transition-duration:, transition-property:, transition-timing-function:,
  }
}

\lstdefinestyle{htmlcssjs} {%
  basicstyle={\footnotesize\ttfamily\small},   
  frame=b,
  xleftmargin={0.75cm},
  numbers=left,
  stepnumber=1,
  firstnumber=1,
  numberfirstline=true,	
  identifierstyle=\color{black},
  keywordstyle=\color{blue}\bfseries,
  ndkeywordstyle=\color{editorGreen}\bfseries,
  stringstyle=\color{editorOcher}\ttfamily,
  commentstyle=\color{editorGreen}\ttfamily,
  language=HTML5,
  alsolanguage=JavaScript,
  alsodigit={.:;},	
  tabsize=2,
  showtabs=false,
  showspaces=false,
  showstringspaces=false,
  extendedchars=true,
  breaklines=true,
  literate=%
  {Ö}{{\"O}}1
  {Ä}{{\"A}}1
  {Ü}{{\"U}}1
  {ß}{{\ss}}1
  {ü}{{\"u}}1
  {ä}{{\"a}}1
  {ö}{{\"o}}1
}
\lstdefinestyle{py} {%
language=python,
literate=%
*{0}{{{\color{lightred}0}}}1
{1}{{{\color{lightred}1}}}1
{2}{{{\color{lightred}2}}}1
{3}{{{\color{lightred}3}}}1
{4}{{{\color{lightred}4}}}1
{5}{{{\color{lightred}5}}}1
{6}{{{\color{lightred}6}}}1
{7}{{{\color{lightred}7}}}1
{8}{{{\color{lightred}8}}}1
{9}{{{\color{lightred}9}}}1,
basicstyle=\footnotesize\ttfamily, 
numbers=left,               
numbersep=5pt,              
tabsize=4,                  
extendedchars=true,         %
breaklines=true,            
keywordstyle=\color{blue}\bfseries,
frame=b,
commentstyle=\color{editorGreen}\itshape,
stringstyle=\color{editorOcher}\ttfamily, 
showspaces=false,           
showtabs=false,             
xleftmargin=17pt,
framexleftmargin=17pt,
framexrightmargin=5pt,
framexbottommargin=4pt,
showstringspaces=false,      
}%

\lstdefinestyle{stack} {%
language=python,
basicstyle=\footnotesize\ttfamily, 
numbers=left,               
numbersep=10pt,              
tabsize=4,                  
extendedchars=true,         %
breaklines=true,            
frame=b,
showspaces=false,           
showtabs=false,             
xleftmargin=17pt,
framexleftmargin=17pt,
framexrightmargin=5pt,
framexbottommargin=4pt,
showstringspaces=false,      
}%
\begin{document}
\graphicspath{{img/}}
\title{An Empirical Study of JavaScript Inclusion Security Issues in Chrome Extensions}
\titlerunning{JISI in Chrome Extensions}
%
%

\author{Chong Guan}
\institute{}
\institute{Zhejiang Gongshang University, China \\
\email{guanchong@zjgsu.edu.cn}}


%
\maketitle              

\newcommand{\extensionNum}{36324}
\newcommand{\extensionUnpubNum}{0}
\newcommand{\extensionPopularNum}{13791}
\newcommand{\extensionGeneralNum}{16224}
\newcommand{\extensionUnusedNum}{8780}
\newcommand{\extensionWithFiveMoreVersion}{62}
\newcommand{\jsJsNum}{723798}
\newcommand{\jsJsFileNum}{178148}
\newcommand{\jsCsNum}{85690}
\newcommand{\jsBsNum}{54514}
\newcommand{\jsIsNum}{583594}
\newcommand{\jsRemoteNum}{29992}
\newcommand{\jsDynamicNum}{207874}
\newcommand{\jsStaticNum}{300459}
\newcommand{\jsSpecStaticNum}{126004}
\newcommand{\jsSpecDynamicNum}{33419}
\newcommand{\jsDyStaIntersectNum}{174455}
\newcommand{\jsLocalhostNum}{27}
\newcommand{\jsipNum}{6}
\newcommand{\jsRingNum}{3}
\newcommand{\jsVulNum}{33}
\newcommand{\jsDomainNum}{9972}
\newcommand{\jsHttpsDomainNum}{534}
\newcommand{\jsHttpDomainNum}{257}
\newcommand{\jsRemoteFileNum}{2231}
\newcommand{\jsStaticRemoteFileNum}{817}
\newcommand{\jsDynamicRemoteFileNum}{2126}
\newcommand{\jsJsPerDomain}{40.37}
\newcommand{\extensionOldNum}{36324}
\newcommand{\libLibraryNum}{10}
\newcommand{\libLibraryIncNum}{79291}
\newcommand{\libLibraryFileNum}{7890}
\newcommand{\libLibraryExtNum}{18138}
\newcommand{\libLibraryPerHost}{6.06}
\newcommand{\newExtNum}{18890}
\newcommand{\newExtUnpubNum}{0}
\newcommand{\newExtPopularNum}{4322}
\newcommand{\newExtGeneralNum}{600}
\newcommand{\newExtUnusedNum}{0}
\newcommand{\newExtWithFiveMoreVersion}{6}
\newcommand{\newJsJsNum}{198121}
\newcommand{\newJsJsFileNum}{47496}
\newcommand{\newJsCsNum}{27670}
\newcommand{\newJsBsNum}{14793}
\newcommand{\newJsIsNum}{155658}
\newcommand{\newJsRemoteNum}{5551}
\newcommand{\newJsDynamicNum}{53699}
\newcommand{\newJsStaticNum}{81631}
\newcommand{\newJsSpecStaticNum}{35563}
\newcommand{\newJsSpecDynamicNum}{7631}
\newcommand{\newJsDyStaIntersectNum}{46068}
\newcommand{\newJsLocalhostNum}{3}
\newcommand{\newJsipNum}{0}
\newcommand{\newJsVulNum}{3}
\newcommand{\newJsDomainNum}{3399}
\newcommand{\newJsHttpsDomainNum}{204}
\newcommand{\newJsHttpDomainNum}{30}
\newcommand{\newJsRemoteFileNum}{783}
\newcommand{\newJsStaticRemoteFileNum}{263}
\newcommand{\newJsDynamicRemoteFileNum}{727}
\newcommand{\newJsJsPerDomain}{25.23}
\newcommand{\newLibraryNum}{10}
\newcommand{\newLibraryIncNum}{9298}
\newcommand{\newLibraryFileNum}{4167}
\newcommand{\newLibraryExtNum}{1216}
\newcommand{\newLibraryPerHost}{2.24}

\begin{abstract}

JavaScript, a scripting language employed to augment the capabilities of web browsers within web pages or browser extensions, utilizes code segments termed JavaScript inclusions. While the security aspects of JavaScript inclusions in web pages have undergone substantial scrutiny, a thorough investigation into the security of such inclusions within browser extensions remains absent, despite the divergent security paradigms governing these environments. This study presents a systematic measurement of JavaScript inclusions in Chrome extensions, employing a hybrid methodology encompassing static and dynamic analysis to identify these inclusions. The analysis of 36,324 extensions revealed 350,784 JavaScript inclusions. Subsequent security assessment indicated that, although the majority of these inclusions originate from local files within the extensions rather than external servers, 22 instances of vulnerable remote JavaScript inclusions were identified. These remote inclusions present potential avenues for malicious actors to execute arbitrary code within the extension's execution context. Furthermore, an analysis of JavaScript library utilization within Chrome extensions disclosed the prevalent use of susceptible and outdated libraries, notably within numerous widely adopted extensions.

\end{abstract}

\section{Introduction}%
\label{sec:introduction}


Chrome extensions are small web applications developed in JavaScript to extend the functionality of the Chrome browser. For example, the AdBlock extension removes advertisements in web pages and has millions of users. Imported third-party JavaScript code has been problematic in web applications, introducing many vulnerabilities~\cite{DBLP:conf/ccs/NikiforakisIKAJKPV12, DBLP:conf/ndss/LauingerCA0WK17}. The risks are amplified in Chrome extensions, as they possess greater privileges. Consequently, a vulnerable or malicious imported JavaScript code within a Chrome extension can lead to more severe consequences.

The security implications of imported third-party JavaScript have been thoroughly investigated in standard web pages. For example, Nikiforakis et al.~\cite{DBLP:conf/ccs/NikiforakisIKAJKPV12} examined JavaScript inclusions in the top 10,000 Alexa~\cite{website:Alexa} websites, evaluated the servers providing the JavaScript, and identified four categories of vulnerabilities associated with insecure third-party inclusion practices. Tobias et al.~\cite{DBLP:conf/ndss/LauingerCA0WK17} analyzed client-side JavaScript library usage and discovered that numerous websites employ outdated and vulnerable libraries.

However, a comprehensive examination of JavaScript inclusion issues in Chrome extensions, which possess a unique security architecture, is lacking. First, JavaScript is utilized in diverse components of a Chrome extension. In web applications, all JavaScript code is within \emph{script} HTML tags. Chrome extensions employ a privilege-separation architecture~\cite{DBLP:conf/uss/ProvosFH03}, segregating an extension into two distinct components: \emph{ContentScript} and the \emph{core extension}. ContentScripts interact with web pages rendered in the Chrome browser but operate without elevated privileges. Core extensions, conversely, do not directly interact with web pages and execute with full privileges. Second, Chrome extensions have more extensive privileges than web applications, managed through a permission system. Extensions must request permissions from the user during installation. Consequently, malicious code injected through vulnerable JavaScript inclusions can only exploit the granted permissions.

This paper reports on an empirical study of JavaScript inclusions in Chrome extensions, undertaken to facilitate an evaluation of Chrome extension security. Initially, a collection of {\extensionNum} Chrome extensions was assembled, and a subset of {\newExtNum} extensions was subjected to longitudinal version tracking over a six-month period. Subsequently, JavaScript inclusions within the Chrome extensions were identified using both dynamic and static analysis. An empirical study of vulnerable JavaScript inclusions revealed {\jsVulNum} instances of remote inclusions susceptible to manipulation, enabling arbitrary code execution within the extension's execution environment. Finally, an analysis of JavaScript library usage within Chrome extensions identified the prevalent use of vulnerable and outdated libraries in these extensions.

The conventional methodology for gathering JavaScript inclusions in web applications proved inadequate for the context of Chrome extensions. Specifically, web application analysis typically employs a proxy to intercept JavaScript inclusion requests between the browser and server. However, this approach is ineffective for Chrome extensions, as local JavaScript file imports do not generate web requests. To address this, we employed a hybrid approach, integrating static and dynamic analysis methods. The static analysis involved parsing web pages and extracting JavaScript inclusions identified by the  {\em script} HTML tag. To capture dynamically generated scripts, which static analysis overlooks, we implemented a dynamic analysis technique. This involved injecting and executing JavaScript code within the loaded extension's web page to enumerate all JavaScript inclusions.

%
%


Static and dynamic methods provide complementary identification of JavaScript inclusions: static analysis detects inclusions before webpage loading, while dynamic analysis captures inclusions in the fully rendered webpage. For instance, JavaScript inclusions within a redirect page, which is not fully rendered, are not detectable by dynamic analysis. In total, we analyzed {\extensionNum} Chrome extensions and identified {\jsJsNum} JavaScript inclusions. Of these, 16.54\% were ContentScripts, 11.32\% were background scripts, and the remainder were embedded within extension webpages. A total of {\jsRemoteNum} inclusions were loaded from external servers.

To investigate vulnerabilities within Chrome extension JavaScript inclusions, we adopted the analytical framework developed by Nikiforakis et al. (2012). Their research, which analyzed JavaScript inclusions in the top 10,000 Alexa websites, delineated four categories of vulnerable JavaScript inclusions.  Based on these categories, our study identified {\jsVulNum} vulnerable JavaScript inclusions in Chrome extensions, which could enable attackers to execute arbitrary code within susceptible extensions.

Compared to the Alexa top 10,000 websites, Chrome extensions exhibit a lower prevalence of vulnerable cases. This disparity arises from the tendency of Chrome extensions to utilize local JavaScript inclusions rather than retrieving them from remote servers.  While 687 distinct remote servers provided JavaScript inclusions to Chrome extensions, Nikiforakis et al. (2012) identified 20,225 unique remote servers.  However, when considering only remote JavaScript inclusions, the proportion of vulnerable instances is approximately equivalent between Chrome extensions and web applications.

Our investigation was expanded to encompass JavaScript libraries employed within Chrome extensions, given that these libraries contain frequently reused code, potentially significantly influencing Chrome extension security.  These libraries may be integrated as local files via copying or from a remote server.  However, the source code of a JavaScript library is not invariably static; even identical versions may vary across different servers, and developers may modify the source code to meet specific requirements.

Consequently, static methods are inadequate for reliable JavaScript library identification.  Therefore, we employed dynamic detection to identify JavaScript libraries within Chrome extensions.  By invoking the version retrieval APIs provided by these libraries when the JavaScript code is loaded in the Chrome browser, we were able to identify the library and its version.  We examined 10 commonly used libraries to determine their presence within the imported JavaScript code.

Our analysis detected \libLibraryIncNum{} library inclusions.  Similar to the distribution of JavaScript inclusions, extension webpages contained the highest number of libraries, with only a small fraction imported from remote servers.  JQuery was found to be the most prevalent library.  Given that the frequency of other library inclusions was substantially lower than that of JQuery, and their analysis yielded similar results, we focus on presenting the JQuery analysis in this paper.  We present the version distribution of JQuery and observe that developers typically update to the latest patch version.  By calculating the age of the libraries relative to the extension's update time, we determined that popular extensions tend to utilize older, outdated libraries.  We also analyzed the prevalence of vulnerable library versions in Chrome extensions, finding that 21.88\% of extensions still employ at least one vulnerable version of the 10 libraries examined.


In summary, the contributions of this study are as follows:

\begin{itemize}
\item We evaluate the performance of static and dynamic JavaScript inclusion detection methodologies on Chrome extensions.
\item We present a detailed analysis of imported JavaScript usage in Chrome extensions, identifying {\jsVulNum} vulnerable JavaScript inclusions based on four vulnerability types.
\item We identify JavaScript libraries within Chrome extensions and determine that a significant proportion of these libraries are outdated, with 21.88\% of extensions employing vulnerable library versions.
\end{itemize}

\section{Background}%
\label{sec:background}

\subsection{JavaScript Inclusion}%
\label{sub:javascript_inclusion}

In web applications, JavaScript inclusions encompass all JavaScript code integrated via the \emph{script} HTML tag.  Examples of JavaScript inclusions in web applications are shown in Listing 1.

The first line illustrates a remote JavaScript inclusion, where the code is retrieved from a service provider's server at the domain \url{www.provider.com}. The second line demonstrates a local JavaScript inclusion. In this instance, the script's src attribute lacks a server address, and the code is loaded from the same server as the webpage. Lines 3-5 provide an example of inline JavaScript, where the src attribute is absent, and the JavaScript code is embedded directly within the script tag.

\begin{figure}
	\begin{lstlisting}[title=Listing 1: JavaScript Inclusions, label=lst:JavaScriptInclusion, style=htmlcssjs]
<script src="https://www.provider.com/sdk.js"></script>
<script src="js/functions.js"></script>
<script>
    console.log("Inline JavaScript demo.")
</script>
    \end{lstlisting}
\end{figure}

The concept of JavaScript inclusion is broadened in Chrome extensions due to their distinct architecture, which incorporates two additional file types,  {\em ContentScript} and {\em background} scripts, that may also contain JavaScript code.  The term ``imported JavaScript" is used here to refer to reused JavaScript code that is either incorporated via direct copying or remote inclusion.

\subsection{Architecture of Chrome Extension}%
\label{sub:chrome_extension_structure}

\begin{figure}[htpb]
    \centering
    \includegraphics[width=0.8\linewidth]{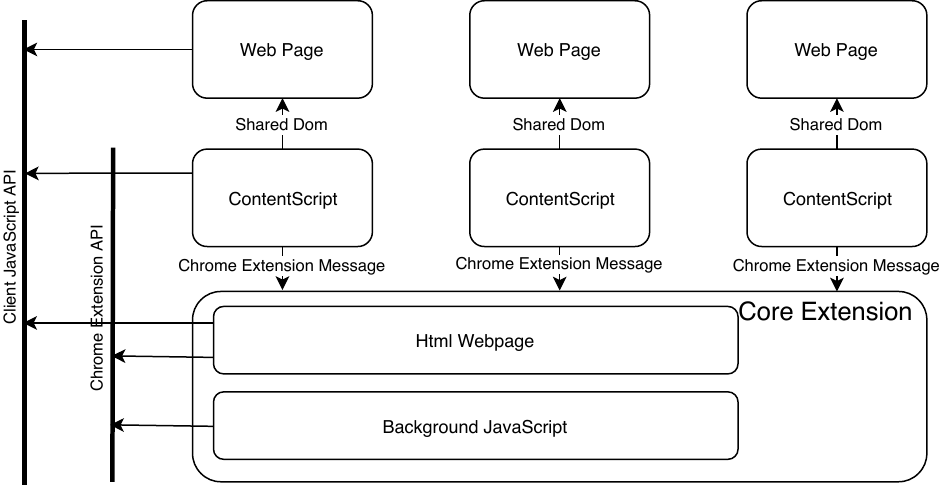}
    \caption{The Architecture of Chrome Extensions}
    \label{fig:architecture}
\end{figure}

Chrome extensions are software applications that operate within the Chrome browser. They enhance the browsing experience, enabling users to customize Chrome's functionality and behavior according to individual requirements or preferences.  For example, AdBlock, a widely used extension, functions to eliminate advertisements from websites such as YouTube or Facebook.  Chrome extensions are developed using web technologies, including HTML, JavaScript, and CSS, and are designed adhering to the principle of least privilege.  All privileges required by a given extension are specified in a manifest file.

Figure \ref{fig:architecture} illustrates the architecture of Chrome extensions, in which JavaScript code can be present in two components with differing privilege levels.  The core extension represents the primary component and possesses all privileges declared in its manifest file.  ContentScripts are JavaScript codes that execute within the context of web pages.  While isolated from the web pages' local JavaScript, ContentScripts share the Document Object Model (DOM) of the webpage.

\subsection{JavaScript Inclusion in Chrome Extension}%
\label{sub:javascript_inclusion_in_chrome_extension}

A Chrome extension comprises ContentScripts and core extension scripts. Core extension scripts can be further categorized into background scripts and scripts within Chrome extension webpages.  Background scripts execute in the extension's background and manage its primary functionality.  Chrome extension webpages provide user interfaces, such as popup pages and option pages.  These webpages can import JavaScript code from local extension files or remote servers.  Generally, background scripts and ContentScripts are local files within the extension.  To enhance functionality and expedite development, developers frequently integrate third-party libraries, either by copying them as local files within the extension or by importing them from a remote server into a Chrome extension webpage.

\section{Data Collection}%
\label{sec:data_collection}
This section delineates the methodologies employed in gathering JavaScript inclusions within Chrome extensions and identifying libraries within those inclusions.

\subsection{Extension Collection}%
\label{sub:downloading_extension}

To acquire Chrome extensions, we conducted an analysis of the source code of the Chrome Web Store~\cite{website:ChromeStore} to extract each extension's unique identifier.  Subsequently, we used these identifiers to construct URLs for downloading the corresponding Chrome extension CRX files.  These CRX files were then decompressed to obtain the source code, which was stored on a server for subsequent analysis.

In addition to the extensions' source code, we also gathered supplementary extension information from the Chrome Web Store, including update timestamps, user counts, rating scores, and supported languages.

\subsection{JavaScript Detection}%
\label{sub:javascript_detection}
As detailed in Section~\ref{sub:javascript_inclusion_in_chrome_extension}, JavaScript code in Chrome extensions can be categorized into three components: background scripts, ContentScripts, and JavaScript within Chrome extension webpages. Background scripts and ContentScripts are consistently local files, and their respective file paths are specified in the extension's manifest file.  By parsing this file, we can efficiently locate ContentScripts and background scripts.  The principal challenge lies in detecting JavaScript within the extension webpages.

We used a combination of static and dynamic methods to address this challenge.  The static method involves parsing webpage content and extracting JavaScript code referenced by \emph{script} tags.  However, this approach cannot detect dynamically generated JavaScript, a feature that allows developers to import JavaScript code at runtime.  Specifically, a \emph{script} element can be created and appended to the webpage during execution.  Dynamically generated JavaScript is prevalent in third-party widgets.  To simplify widget integration, developers often provide a single JavaScript file for inclusion via the \emph{script} HTML tag, which then generates other required JavaScript inclusions dynamically.

Request interception using a proxy represents one technique for detecting dynamically generated JavaScript.  However, this method is unsuitable for Chrome extensions, because importing a local JavaScript file from an extension webpage does not initiate a network request.

To capture dynamically generated JavaScript, we developed a dynamic method.  This involves loading extensions and opening target webpages in the Chrome browser using Selenium.  We then inject the following JavaScript code:
\emph{document.getElementByTagName("script")}
into the extension webpage to enumerate all JavaScript inclusions within the loaded page.

\section{Analysis Results}%
\label{sec:analysis}
This section delineates the outcomes of our investigation. Initially, we present descriptive statistics concerning the assembled corpus of Chrome extensions. Subsequently, we detail the analysis of JavaScript inclusions and the findings from vulnerability assessments targeting remote server JavaScript inclusions. Finally, we examine the utilization of prevalent JavaScript libraries within Chrome extensions.

\subsection{Chrome Extensions}%
\label{sub:chrome_extensions}

\begin{figure*}[t!]
    \centering
\subfloat[Extension Language]{\includegraphics[width=0.3\textwidth]{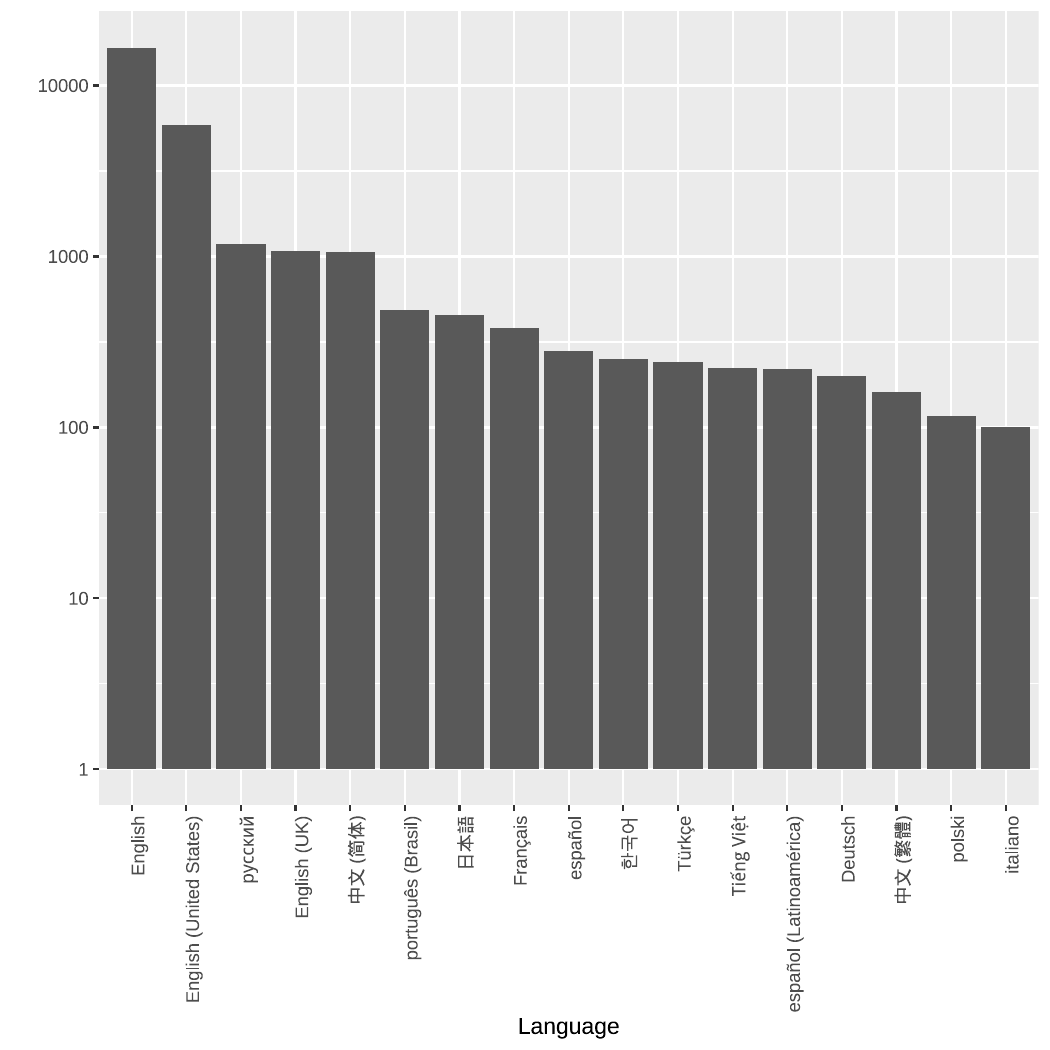}
\label{fig:language}}
\hfil
\subfloat[User Number]{\includegraphics[width=0.3\textwidth]{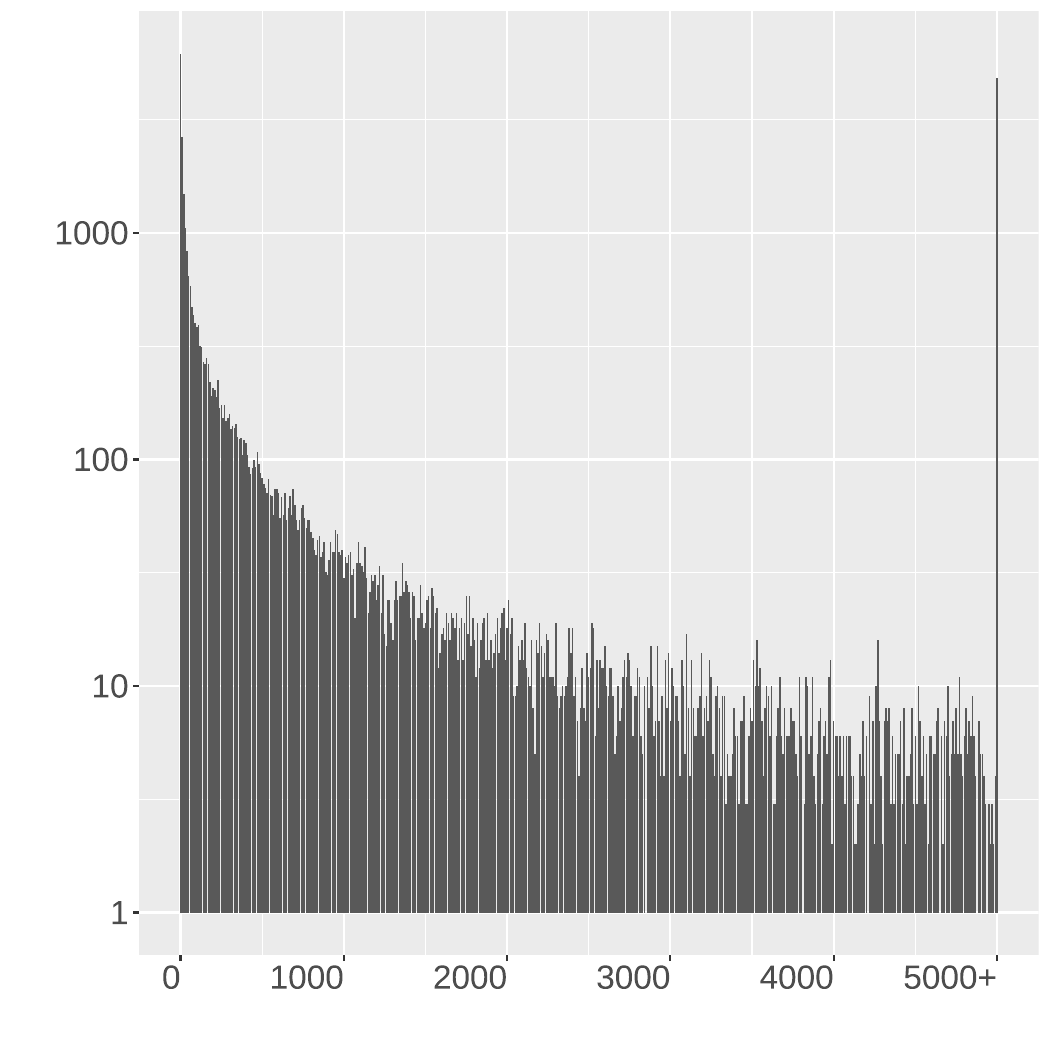}
\label{fig:userNum}}
\hfil
\subfloat[Rated Score]{\includegraphics[width=0.3\textwidth]{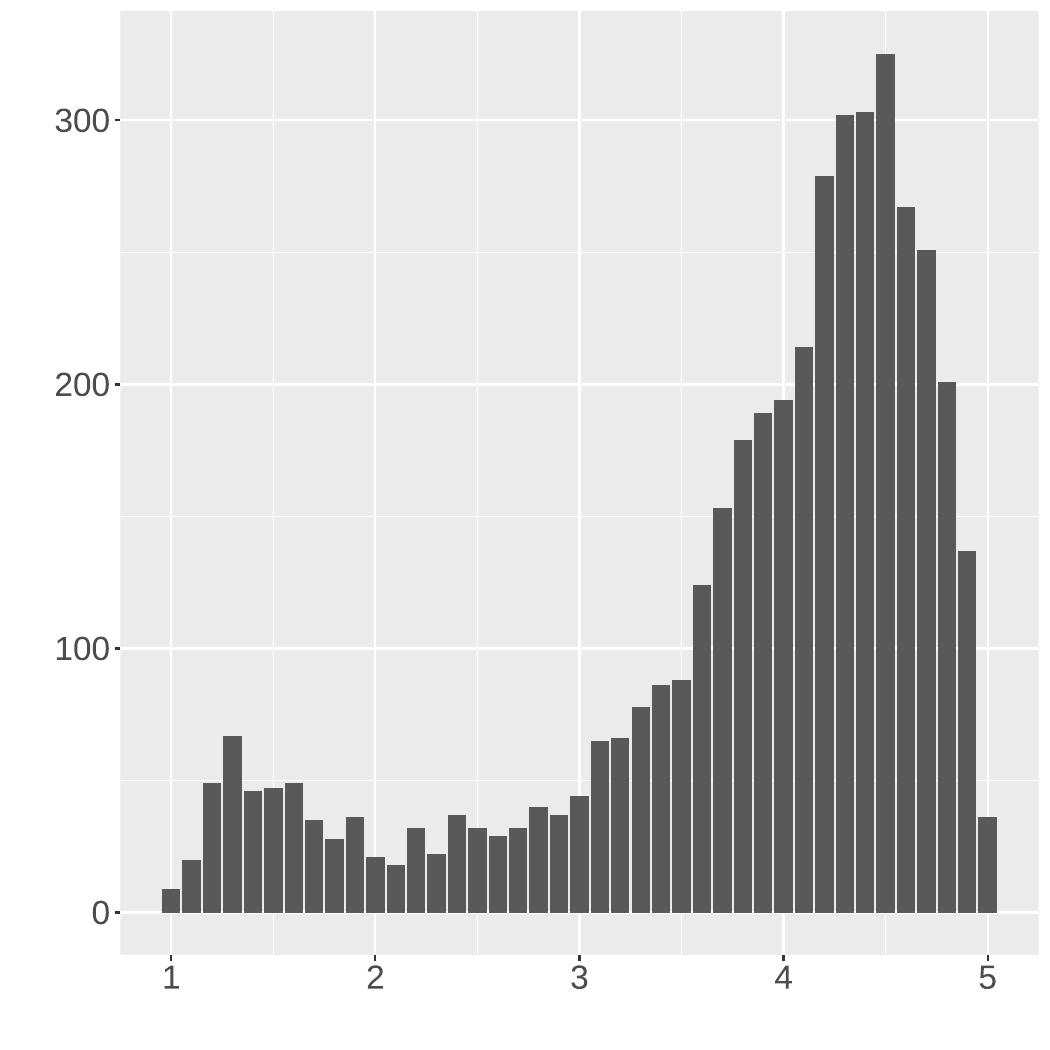}
\label{fig:ratedScore}}
\newline
\subfloat[Updatetime]{\includegraphics[width=0.3\textwidth]{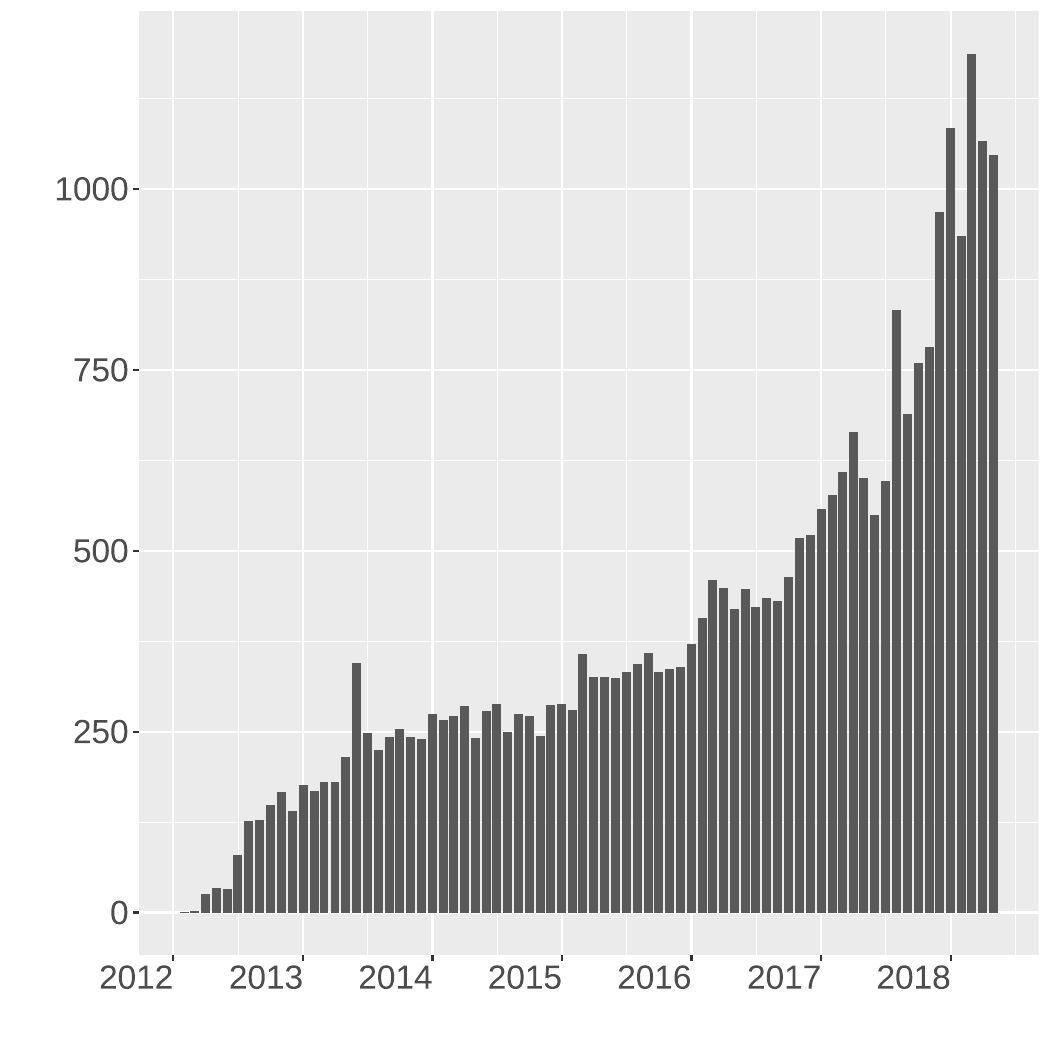}
\label{fig:updatetime}}
\subfloat[Version]{\includegraphics[width=0.3\textwidth]{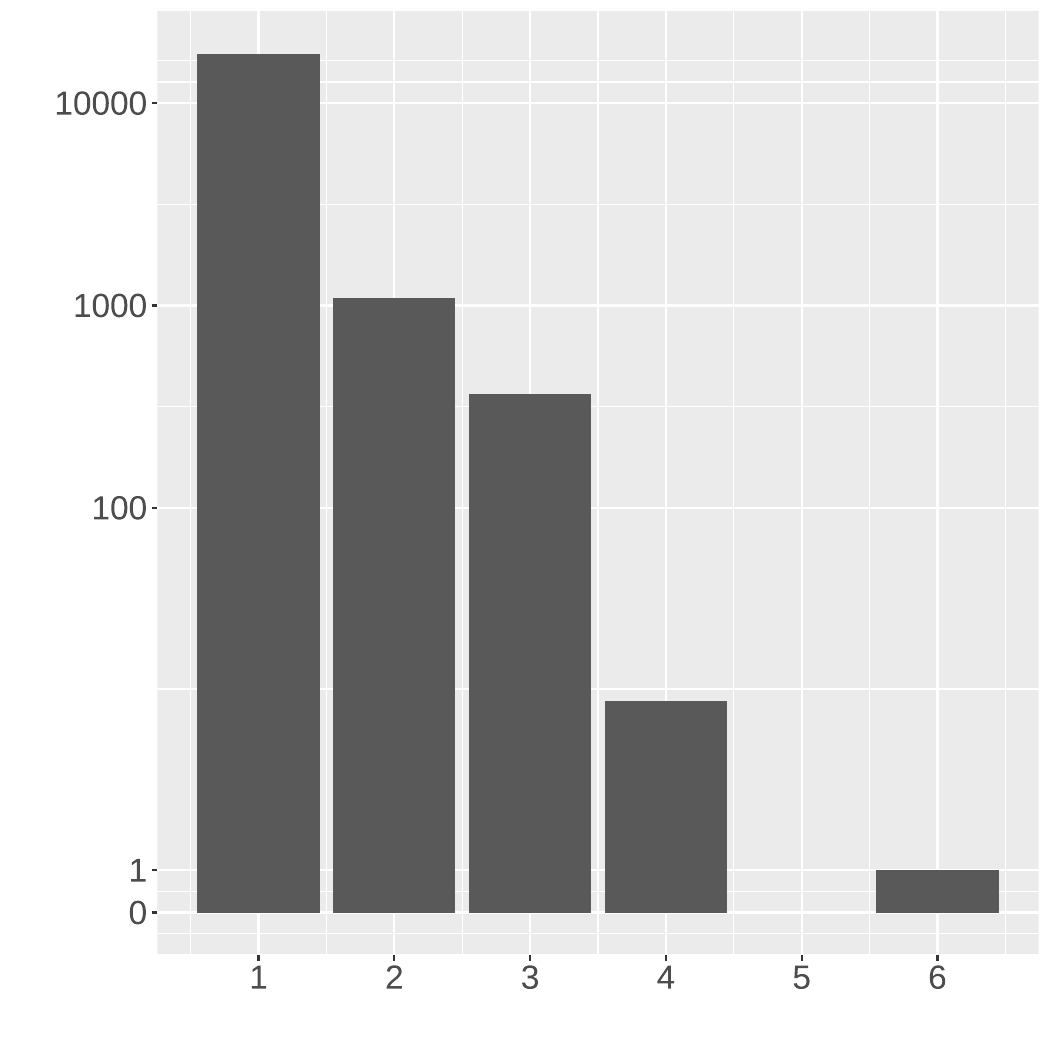}
\label{fig:version}}
\caption{The Count of Extensions with Different Attributes}
\label{fig:ext_dist}
\end{figure*}

Our collection comprises a total of {\extensionNum} extensions sourced from the Chrome Web Store. The initial dataset was compiled at the close of 2018, and from August 2021 to February 2022, we have monitored {\newExtNum} extensions, acquiring multiple versions thereof. Figure~\ref{fig:ext_dist} illustrates the distribution of these extensions based on their programming language, user count, rated score, last update timestamp, and version. Figure~\ref{fig:language} displays the extension count for the top 17 languages, revealing that the majority of Chrome extensions cater to English-speaking users. Figure~\ref{fig:userNum} indicates that only a small fraction of extensions are utilized by a substantial user base. Based on this observation, we categorize the dataset into three groups for subsequent analysis: popular extensions (over 1000 users, {\extensionPopularNum}), general extensions (20 to 1000 users, {\extensionGeneralNum}), and infrequently used extensions (0 to 20 users, {\extensionUnusedNum}). Figure~\ref{fig:updatetime} presents the last modification date for all original extension samples collected in 2018, revealing a significant number of extensions that are not actively maintained. For {\newExtNum} extensions, we obtained newer iterations in 2021, and Figure~\ref{fig:version} shows the distribution of extensions according to the number of versions available.

\subsection{General JavaScript Statistics}%
\label{sub:general_javascript_statistics}

\begin{table*}[t]
    \centering
\begin{tabular}{lccccc}
\toprule[1pt]

\multirow{3}{*}{\begin{tabular}[c]{@{}l@{}}Extension\\Type\end{tabular}} & \multirow{3}{*}{\begin{tabular}[c]{@{}l@{}}Total\\Number\end{tabular}} & \multicolumn{4}{c}{JavaScript Inclusion(File)Frequency Per Extension}                              \\  \cline{3-6}
                            &       & \multicolumn{2}{c|}{Local}& \multicolumn{1}{c|}{Local} & Remote                     \\ \cline{3-6}
                            &       & \multicolumn{1}{c|} {BackgroundScript}        & ContentScript        & \multicolumn{2}{|c}{Script In Extension Webpage} \\ \hline

Popular  & 9769&2.551(1.566)&4.112(2.263)&25.647(5.075)&0.863(0.058) \\
General  & 15869&0.957(0.761)&1.623(1.082)&11.654(2.594)&0.926(0.071) \\
Unused   & 8780&0.763(0.595)&1.117(0.81)&7.079(1.886)&0.338(0.072) \\
Sum      & 34418&1.36(0.947)&2.2(1.347)&14.459(3.117)&0.758(0.067) \\

\bottomrule[1pt]
\end{tabular}
\captionsetup{justification=centering}
\caption{JavaScript Number in Chrome Extension}
\label{tab:js_inclusion_num}
\end{table*}

In total, we identified {\jsJsNum} JavaScript inclusions, comprising {\jsJsFileNum} unique JavaScript files. Table~\ref{tab:js_inclusion_num} presents a summary of the frequencies of various types of JavaScript inclusions (note that the total number is less than {\extensionNum} due to the absence of user count data for some extensions). The majority of JavaScript code is incorporated into extension webpages. Popular extensions tend to exhibit a higher number of JavaScript inclusions. While JavaScript sourced from remote servers constitutes a minor proportion of all inclusions, these files demonstrate frequent reuse across Chrome extensions. Specifically, remotely sourced JavaScript files are reused approximately seven times on average, whereas other JavaScript files are reused no more than twice. Consequently, the compromise of a single remotely included JavaScript file could potentially affect a larger number of extensions compared to other inclusion types.

We detected {\jsRemoteFileNum} JavaScript files originating from remote servers, hosted across {\jsDomainNum} distinct domains, with 63\% of these hosts employing the HTTPS protocol. Table~\ref{tab:top_hosts} lists the top 5 most frequently encountered hosts, with Google's servers being the most prevalent source of JavaScript. On average, each host serves {\jsJsPerDomain} JavaScript inclusions.

\begin{table}[t]
    \centering
    \begin{tabular}{cc}
        \hline
\toprule[1pt]
Hostname            & Extension Number\\ 
 \hline
https://ssl.google-analytics.com/& 1704  \\ 
https://www.google-analytics.com/& 1072  \\ 
https://www.google.com/& 436   \\ 
https://ajax.googleapis.com/& 363   \\ 
https://www.happyhey.com/& 212   \\ 
\bottomrule[1pt]
    \end{tabular}
    \caption{Top Hosts in Remote JavaScript Inclusions}
    \label{tab:top_hosts}
\end{table}

\begin{figure}[t]
    \centering
    \includegraphics[width=0.6\linewidth]{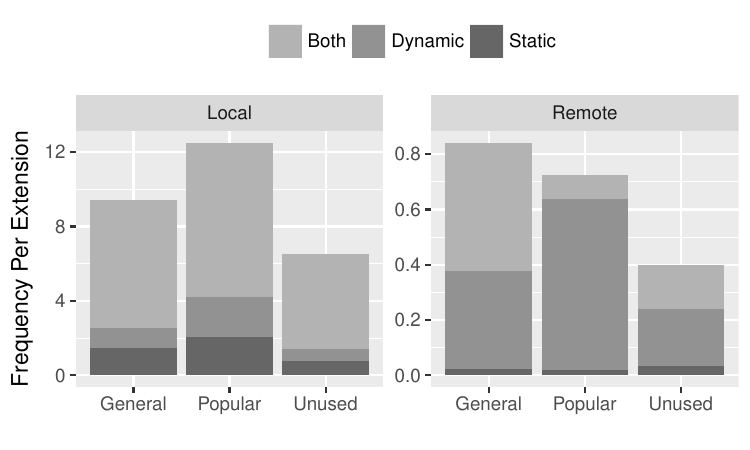}
    \caption{JavaScript Inclusion Frequency in Dynamic and Static Methods}
    \label{fig:jsPlot/dsComparison}
\end{figure}

Figure~\ref{fig:jsPlot/dsComparison} presents a bar chart comparing the frequency of JavaScript inclusions identified through static and dynamic analysis. Surprisingly, static analysis detected a greater number of local JavaScript inclusions than dynamic analysis. Further investigation revealed that this discrepancy arises from the dynamic method missing certain transient JavaScript inclusions. The dynamic method captures all JavaScript inclusions present after webpage loading; however, some inclusions are employed temporarily and subsequently removed immediately. For instance, some webpages serve as redirects and are not rendered as final content. JavaScript inclusions within these pages are consequently missed by the dynamic method. Generally, both static and dynamic approaches are effective in detecting the majority of JavaScript inclusions. We found {\jsDyStaIntersectNum} JavaScript files identified by both methodologies, {\jsSpecDynamicNum} exclusively by the dynamic method, and {\jsSpecStaticNum} solely by the static method. Nevertheless, the dynamic method significantly outperformed the static method in identifying remote JavaScript inclusions, as the dynamic creation of JavaScript inclusions is typically associated with importing remote scripts within JavaScript libraries, rather than local ones.

\subsection{Vulnerabilities in Remote JavaScript Inclusion}%
\label{sub:attack_analysis}

Improperly configured remote JavaScript inclusions can enable attackers to execute arbitrary code. We evaluated four categories of vulnerabilities and compared the results with those reported for the Alexa top 10,000 websites~\cite{DBLP:conf/ccs/NikiforakisIKAJKPV12}.

The first vulnerability pertains to cross-user and cross-network scripting. If a remote address is specified as localhost or 127.0.0.1, the JavaScript inclusion will attempt to retrieve code from the user's local machine. In multi-user environments, an attacker can listen on a port and serve malicious JavaScript code without requiring root privileges if the remote address includes a port number greater than 1024. If the remote address specifies a private IP address (e.g., 192.168.2.2), the JavaScript will be retrieved from the user's local network, potentially allowing its control by an unauthorized entity. Within our dataset, we identified {\jsRingNum} JavaScript inclusions with the 127.0.0.1 address and 16 instances using localhost. In contrast, the Alexa top 10,000 websites exhibited a higher prevalence of such vulnerable inclusions (131 among the 10,000 websites)~\cite{DBLP:conf/ccs/NikiforakisIKAJKPV12}.

The second vulnerability involves stale Domain-Name-Based Inclusions. A domain name may expire, and if the original registrant does not renew it, an attacker can acquire the domain and serve malicious content. Research on the Alexa top 10,000 websites revealed 56 such domains~\cite{DBLP:conf/ccs/NikiforakisIKAJKPV12}. In Chrome extensions, we identified 2 cases of this vulnerability.

\begin{table*}[ht!]
\centering
\begin{tabular}{lcccc}
\toprule[1pt]

\multirow{3}{*}{Library} & \multicolumn{4}{c}{Library JavaScript Inclusion (Extension)  Frequency}                              \\  \cline{2-5}
                                   & \multicolumn{2}{c|}{Local}& \multicolumn{1}{c|}{Local} & Remote                     \\ \cline{2-5}
                                   & \multicolumn{1}{c|} {BackgroundScript}        & ContentScript        & \multicolumn{2}{|c}{JavaScript In Extension Webpage} \\ \hline

jquery     & 3888(3869) & 9342(7423) & 24433(11742) & 1296(770) \\
Underscore & 294(291)   & 584(452)   & 1414(827)    & 15(10)    \\
Moment     & 213(213)   & 321(266)   & 972(632)     & 16(13)    \\
RequireJS  & 51(51)     & 53(47)     & 1567(446)    & 6(5)      \\
Backbone   & 15(15)     & 55(31)     & 162(113)     & 1(1)      \\
Mustache   & 7(7)       & 68(44)     & 173(124)     & 0(0)      \\
Handlebars & 5(5)       & 64(60)     & 235(161)     & 4(4)      \\
Modernizr  & 4(4)       & 11(11)     & 219(117)     & 14(14)    \\
Knockout   & 3(3)       & 8(8)       & 233(191)     & 10(10)    \\

\bottomrule[1pt]
\end{tabular}
\caption{Library Numbers in Chrome Extension}
\label{tab:libDisInsideExtension}
\end{table*}

The third vulnerability concerns stale IP-Address-Based Inclusions. When a web developer uses an IP address as the source for a remote JavaScript inclusion, attackers who gain control of that IP address can serve malicious code to users. Our experimental data revealed 1 instance of this vulnerability, compared to 35 cases found in the Alexa top 10,000 websites~\cite{DBLP:conf/ccs/NikiforakisIKAJKPV12}.

The final vulnerability is Typosquatting Cross-Site Scripting. Typosquatting~\cite{DBLP:conf/fc/MooreE10, DBLP:conf/sruti/WangBWVD06} involves registering domain names that are slight variations of popular website domains. This technique can redirect users who misspell a web address to a malicious site. Similarly, if a developer makes a typographical error in the source attribute of a JavaScript inclusion, the browser might inadvertently request JavaScript code from a malicious domain. We did not detect any such vulnerabilities in our experiment, while 6 cases were reported in the Alexa top 10,000 websites. All identified vulnerable JavaScript inclusions were associated with 15 distinct Chrome extensions, with user counts ranging from 8 to 2030.

In Chrome extensions, the necessary permissions are specified in the manifest file, indicating the types of resources the extension can access and the addresses the extension can interact with. Malicious JavaScript code can exploit these granted permissions to attack users. The 15 vulnerable extensions requested 15 different resource permissions. Table~\ref{tab:permissions} lists these permissions and their frequency. The most frequently requested permissions are \emph{storage} and \emph{tabs}. We identified 22 distinct web server addresses within the permissions of vulnerable extensions, with 8 vulnerable extensions requesting permissions with wildcard web addresses (e.g., http:///).

Our findings suggest that Chrome extensions experience fewer attacks of the types analyzed compared to the Alexa top websites. A primary reason for this is the tendency of Chrome extensions to utilize more local JavaScript rather than fetching it from remote servers. Our analysis identified only {\jsDomainNum} unique remote hosts, whereas 20,225 unique remote hosts were identified in~\cite{DBLP:conf/ccs/NikiforakisIKAJKPV12}. Additionally, we found 3395 JavaScript inclusions attempting to retrieve code from non-existent addresses, both locally and remotely. If an attacker were able to upload files to such addresses, the corresponding Chrome extensions would become vulnerable.

\begin{table*}[t]
    \begin{tabular}{llp{7cm}}
        \toprule[1pt]
Permission         & Frequency & Description \\ \hline
storage            & 11        & To store, retrieve, and track changes to user data. \\
tabs               & 10        & Gives your extension access to privileged fields of the Tab objects.             \\
activeTab          & 6        & Gives an extension temporary access to the currently active tab.             \\
contextMenus       & 4         & To add items to Google Chrome's context menu.             \\
background         & 3         & Makes Chrome start up early and shut down late, so that apps and extensions can have a longer life.             \\
notifications      & 3         & To create rich notifications using templates and show these notifications to users in the system tray.             \\
desktopCapture     & 2         & To capture content of screen, individual windows or tabs.            \\
downloads          & 1         &  To programmatically initiate, monitor, manipulate, and search for downloads.             \\
webRequest         & 1         &  To observe and analyze traffic and to intercept, block, or modify requests in-flight.             \\
webRequestBlocking & 1         & Required if the extension uses the chrome.webRequest API in a blocking fashion.             \\
history            & 1         & To interact with the browser's record of visited pages.             \\
browsingData       & 1         & To remove browsing data from a user's local profile.             \\
webNavigation      & 2         & To receive notifications about the status of navigation requests in-flight.             \\
cookies            & 1         & To query and modify cookies, and to be notified when they change.              \\
unlimitedStorage   & 1         & Provides an unlimited quota for storing HTML5 client-side data. \\
\bottomrule[1pt]
\end{tabular}
\caption{Permissions In Vulnerable Extensions}
\label{tab:permissions}
\end{table*}

\subsection{Library Analysis}%
\label{sub:library_analysis}

\subsection{Library Identification}%
\label{sub:library_identification}
To acquire more granular metadata regarding the JavaScript inclusions, we performed an identification process for prevalent JavaScript libraries within the collected JavaScript files from the Chrome extensions. The identification methodology employed is analogous to the dynamic method described in Section~\ref{sub:javascript_detection}. Most widely adopted libraries provide programmatic interfaces for retrieving their version information. By invoking these APIs, we can ascertain the specific library based on the returned value.

In a prior study~\cite{DBLP:conf/ndss/LauingerCA0WK17}, a collection of 72 popular JavaScript libraries was compiled utilizing popularity metrics from the JavaScript package manager Bower~\cite{website:Bower}, the web technology survey Wappalyzer~\cite{website:Wappalyzer}, and public Content Delivery Networks (CDNs) operated by Google, Microsoft, and Yandex. However, given that certain prominent libraries did not incorporate a version retrieval API in their earlier releases, precise identification of these libraries is not feasible. To enhance the accuracy of our identification process, we selected 10 widely used libraries from~\cite{DBLP:conf/ndss/LauingerCA0WK17} that consistently provide a version retrieval API across all their released versions. These libraries are as follows: jQuery, Underscore, Backbone, Handlebars, Knockout, Modernizr, Moment, Mootools, Mustache, and RequireJS.

In total, we detected {\libLibraryIncNum} JavaScript library inclusions, comprising {\libLibraryFileNum} unique JavaScript files. {\libLibraryExtNum} extensions incorporate at least one of the 10 selected JavaScript libraries. Notably, JavaScript files belonging to these libraries exhibit a higher rate of reuse compared to general JavaScript files (11.45 vs. 2.42). Table~\ref{tab:libDisInsideExtension} presents a summary of the frequency of different libraries within Chrome extensions. The distribution observed is similar to that of general JavaScript inclusions. The ContentScript component contains a greater number of JavaScript libraries compared to background scripts. The majority of JavaScript libraries are utilized within the extension webpages.

The dynamic method demonstrates efficacy in detecting JavaScript libraries imported from remote servers. For RequireJS and Knockout libraries, the dynamic method identified a greater number of local instances compared to the static method. However, upon closer examination of these JavaScript inclusions, we determined that this discrepancy is attributable to the repeated utilization of specific popular components. For instance, a jquery-ui package installed via npm includes numerous HTML webpages that dynamically import requireJS for testing purposes.

Table~\ref{tab:top_hosts_library} lists the most frequently encountered hosts serving the detected JavaScript libraries. Unsurprisingly, Google's CDN is the most extensively utilized. On average, each host serves {\libLibraryPerHost} library inclusions. jQuery emerges as the most frequently employed library within Chrome extensions. Given the relatively small dataset for other libraries and the similarity of their analysis results to those of jQuery, we will focus on presenting the analytical findings for jQuery in the subsequent sections.

\begin{figure*}[t]
    \centering
    \includegraphics[width=1\linewidth]{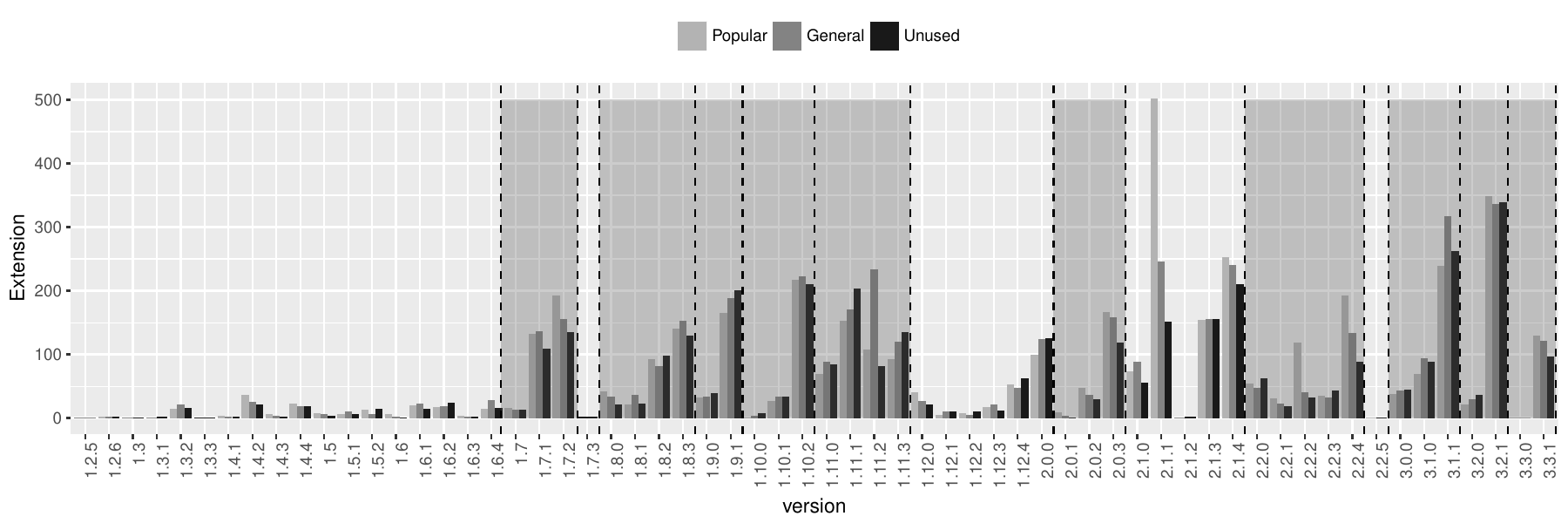}
    \caption{Jquery Version Distribution}
    \label{fig:img/jquery}
\end{figure*}

\begin{table}[t]
    \centering
    \begin{tabular}{cc}
        \hline
\toprule[1pt]
Hostname            & Extension Number\\ 
 \hline
https://ajax.googleapis.com/& 307   \\ 
http://ajax.googleapis.com/& 88    \\ 
http://beauteousbox.com/& 81    \\ 
https://code.jquery.com/& 77    \\ 
http://code.jquery.com/& 53    \\ 

\bottomrule[1pt]
    \end{tabular}
    \caption{Top Hosts in Remote JavaScript Library Inclusion}
    \label{tab:top_hosts_library}
\end{table}

\subsubsection{Jquery Versions In Chrome Extension}%
\label{ssub:jquery_versions_in_chrome_extension}
To gain a deeper understanding of JavaScript library utilization within Chrome extensions, we extracted the version information for these libraries. Figure~\ref{fig:img/jquery} illustrates the distributional characteristics of the jQuery library versions encountered. The data indicates that jQuery versions prior to 1.7 are infrequently used, more recent versions tend to exhibit higher usage frequencies, and older versions of jQuery remain prevalent. The version naming convention adheres to the major.minor.patch scheme. Our observations suggest that developers typically update to the latest patch version within a given minor version. The two exceptions observed in Figure~\ref{fig:img/jquery} (1.7.3, 2.2.5) do not correspond to officially released jQuery versions. These instances represent jQuery builds modified by developers who have assigned custom version numbers, hence their limited occurrence. We determined that the versions are not ordered chronologically by their release dates. The final few versions within the major version 1 series are maintained concurrently with the major version 2 series by the jQuery team, which likely accounts for the lower adoption rate of version 1.12.* jQuery.

We identified 715 extensions that import multiple versions of jQuery, with the majority of these instances involving outdated jQuery versions. Analysis of the extensions' source code revealed that library dependencies are a primary cause of this multi-version inclusion. Specifically, certain JavaScript libraries depend on specific jQuery versions, leading to the introduction of new jQuery versions into extensions even if a version of jQuery is already present.

\subsubsection{Ages of Jquery Library}%
\label{ssub:ages_of_jquery_library}

\begin{figure}[t!]
    \centering
    \subfloat[Different User Number]{\includegraphics[width=0.45\linewidth]{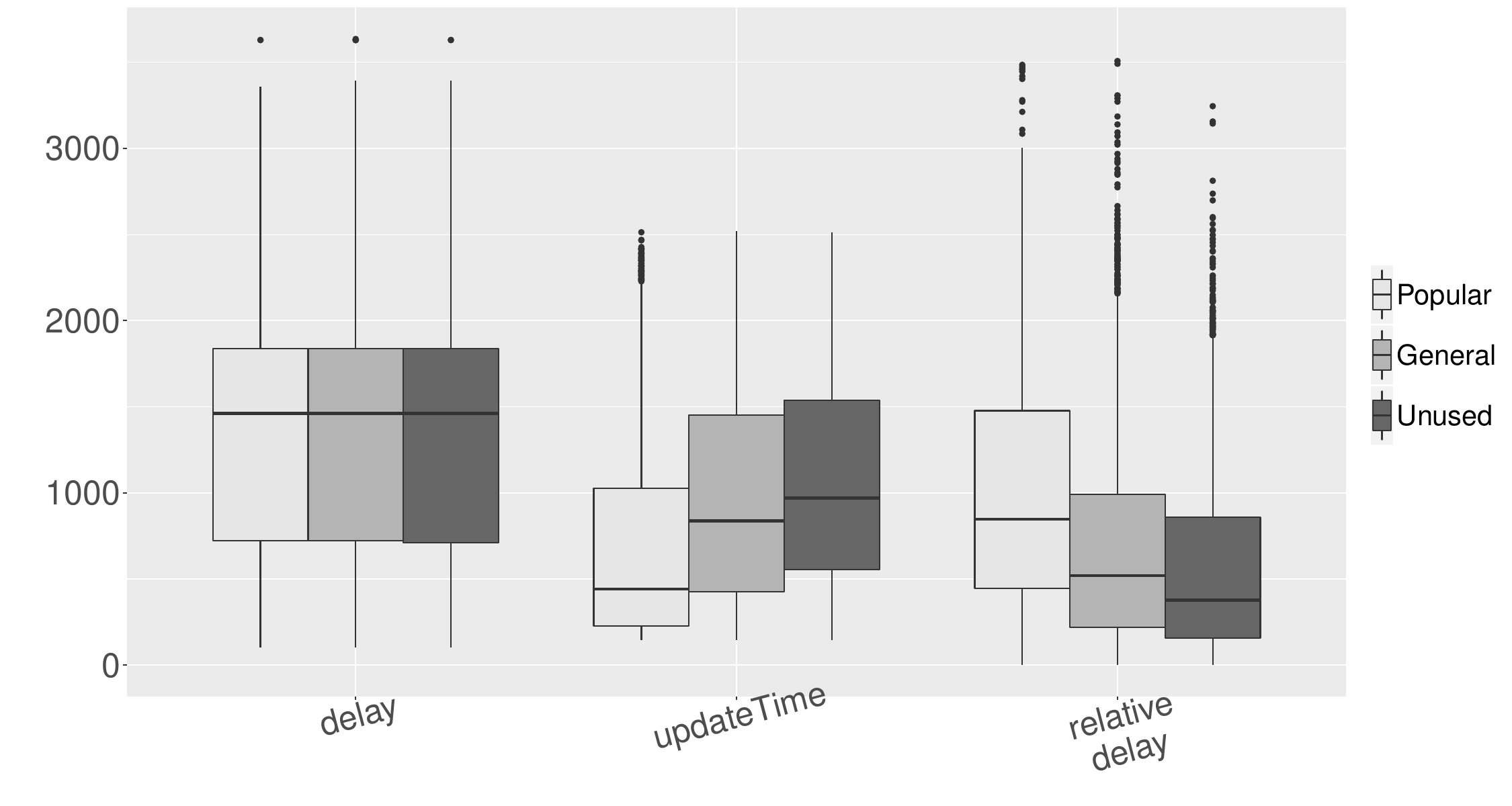}}
    \label{fig:libPlot/raw/lagsByUsernum}
    \hfil
    \subfloat[Different Language]{\includegraphics[width=0.45\linewidth]{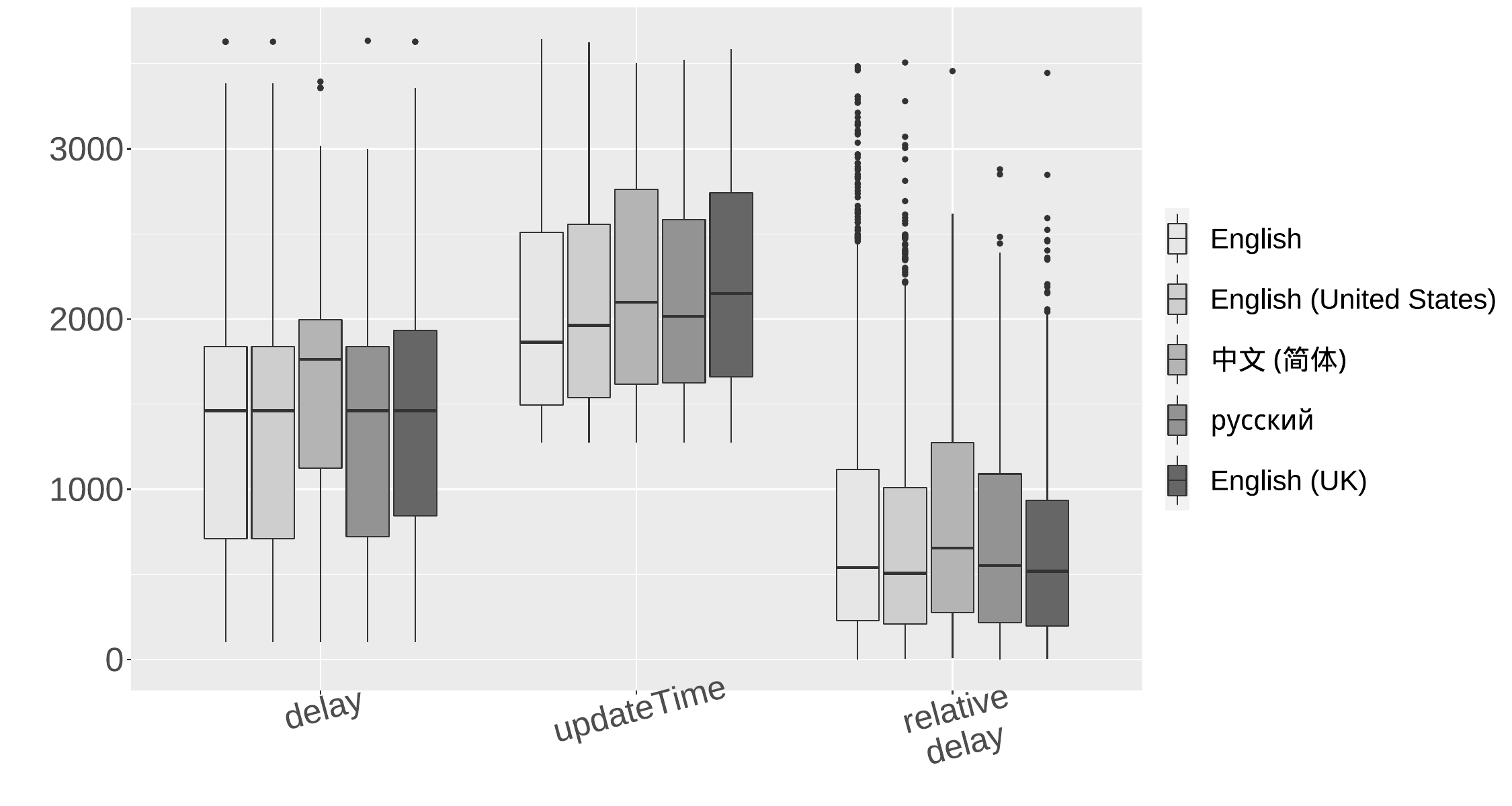}}
    \label{fig:libPlot/raw/lagsByLanguage}
    \caption{The lag days of JavaScript Libraries in Chrome Extensions}
    \label{fig:libPlot/raw/summary}
\end{figure}

To quantify the temporal disparity between the outdated libraries and the most current versions, we retrieved the release dates of jQuery versions by examining the commit history of its GitHub open-source project. We then compared the release dates of the jQuery versions used in the extensions with the data collection date and the extensions' last update dates. Figure~\ref{fig:libPlot/raw/summary} presents the temporal lag of the jQuery libraries in days. The lower and upper hinges correspond to the first and third quartiles (the 25th and 75th percentiles, respectively). The first cluster represents the lag in days of jQuery libraries within Chrome extensions. The second cluster illustrates the temporal difference between the data collection date and the extensions' last update time. The final cluster depicts the relative temporal delay of jQuery libraries, specifically the lag between the jQuery version release time and the extensions' last update time. The temporal lags of jQuery in the majority of extensions exceed 1000 days. Comparing these results with those reported in~\cite{DBLP:conf/ndss/LauingerCA0WK17}, a greater prevalence of outdated libraries is observed in Chrome extensions. While popular extensions tend to undergo regular updates, these updates often do not encompass the associated jQuery libraries. Surprisingly, popular extensions with the most recent update times tend to utilize more outdated jQuery libraries. Furthermore, our analysis based on extension languages indicates that Chinese extensions exhibit a tendency to employ more outdated jQuery libraries.

\subsubsection{Vulnerable Libraries in Chrome Extensions}%
\label{ssub:vulnerable_jquery_in_chrome_extension}
We also investigated the presence of vulnerable libraries within Chrome extensions. The identification of vulnerable libraries was based on information obtained from the npm registry~\cite{website:npm}. Any jQuery version known to contain security vulnerabilities was flagged as such. Our analysis revealed that 5 libraries among the 10 selected libraries contain vulnerable versions. Ultimately, we found that 21.88\% of the analyzed Chrome extensions import at least one vulnerable library.

\section{Discussion}%
\label{sec:discussion}

Despite stringent security protocols, our findings reveal the persistent presence of vulnerable JavaScript inclusions within Chrome extensions, alongside the widespread utilization of deprecated JavaScript libraries. The subsequent sections delineate potential strategies for enhancing the security of JavaScript implementation in Chrome extensions and address the limitations of the current study, outlining avenues for future research.

\subsection{Remediation}%
\label{sub:remediation}

Firstly, the JavaScript inclusion detection methodology developed in this study can be integrated into the security assessment framework for Chrome extensions. The automated acquisition of comprehensive metadata pertaining to JavaScript inclusions facilitates the detection of the four vulnerability classes identified herein. This data can further aid in Google's evaluation of the security risks associated with JavaScript inclusions within Chrome extensions. Furthermore, the automated identification of a significant proportion of outdated JavaScript libraries enables the implementation of developer notifications, alerting them to the potential risks associated with the use of deprecated software components.

Secondly, it is proposed that Google consider the removal of extensions exhibiting low user engagement and a prolonged absence of updates from the Chrome Web Store. Our empirical evidence suggests that while Chrome extension developers generally adopt contemporary JavaScript libraries, the lack of incentive for developers to remove abandoned extensions results in the continued availability of potentially vulnerable software.

\subsection{Limitation and Future Work}%
\label{sub:limitation_and_future_work}
As indicated in Section~\ref{sub:general_javascript_statistics}, the current JavaScript inclusion detection method exhibits certain limitations, resulting in the omission of some inclusions. Specifically, when ContentScripts within Chrome extensions support dynamic JavaScript inclusions, our methodology is unable to detect this category of ContentScript. This is attributed to the injection mechanism relying on Chrome-specific extension APIs. Future work will focus on identifying alternative approaches to comprehensively enumerate all ContentScripts.

Although the data acquisition framework was initially designed for Chrome extensions, the underlying methodology possesses the potential for broader application to the extension ecosystems of other prevalent browsers. Given the architectural similarities between Chrome extensions and those of popular browsers such as Firefox and Safari, and the cross-browser compatibility of the Selenium web automation framework, future research will involve extending our data collection capabilities to encompass extensions from other major browsers. This will enable a comprehensive analysis of JavaScript inclusion usage across the entire spectrum of major browser extensions.

\section{Related Work}%
\label{sec:related_work}

As a foundational language for web development, JavaScript is ubiquitously present within the web environment. The security challenges arising from JavaScript inclusion have been a subject of scholarly inquiry for numerous years. An early empirical investigation into insecure JavaScript practices on the web was conducted by Yue and Wang in 2009~\cite{DBLP:conf/www/YueW09}. Their study utilized a corpus of 6,805 homepages from popular websites to examine risky behaviors associated with JavaScript inclusion and dynamic code generation.
Nikiforakis et al.~\cite{DBLP:conf/ccs/NikiforakisIKAJKPV12} reported on a comprehensive crawl of over three million pages from the top 10,000 Alexa-ranked sites in 2012. Their research identified the provenance of JavaScript inclusions and formulated host-centric metrics for assessing the maintenance quality of remote code providers. They further identified four classes of vulnerabilities potentially exploitable by adversaries to compromise popular websites. Our work evaluates these vulnerabilities within the context of Chrome extensions and compares the empirical findings with the results reported by Nikiforakis et al.

Lauinger et al.~\cite{DBLP:conf/ndss/LauingerCA0WK17} investigated the more specific yet semantically dense domain of libraries. Their study examined the incorporation of deprecated libraries and ascertained the agents and underlying factors contributing to these inclusions. Richards et al.~\cite{DBLP:conf/pldi/RichardsLBV10} and Ratanaworabhan et al.~\cite{DBLP:conf/webapps/RatanaworabhanLZ10} analyzed the dynamic behavior of prevalent JavaScript libraries. Derr et al.~\cite{DBLP:conf/ccs/DerrBFA017} explored the updatability of third-party libraries on the Android platform. Ikram et al.~\cite{DBLP:conf/www/IkramMTKLE19} conducted a large-scale study of dependency chains on the web, revealing that approximately 50\% of first-party websites render content that was not directly loaded by them. Squarcina et al.~\cite{DBLP:conf/uss/SquarcinaTVCM21} investigated the concept of related-domain attackers and performed a measurement study on 50,000 domains. Silva et al.~\cite{DBLP:conf/uss/SilvaNEKYK21} analyzed over eight hundred million VirusTotal (VT) URLs and developed the first content-agnostic machine learning models to differentiate between various types of apex domains hosting malicious websites. Our research aligns with the study of insecure remote JavaScript inclusions but focuses specifically on the domain of Chrome extensions.

Google has developed and implemented a dedicated security architecture for Chrome extensions~\cite{DBLP:conf/uss/KapravelosGCKVP14, DBLP:conf/uss/JagpalDGMPRT15}. Within this security framework and threat model, Carlini et al.~\cite{DBLP:conf/uss/CarliniFW12} conducted a security assessment of 100 Chrome extensions, identifying 70 vulnerabilities across 40 extensions. Liu et al.~\cite{DBLP:conf/ndss/LiuZYC12} demonstrated that Chrome's extension security model does not constitute a universal solution for all potential attacks targeting browser extensions. Malicious extensions continue to represent significant security risks. Starov et al.~\cite{DBLP:conf/sp/StarovN17} and Sj{\"{o}}sten et al.~\cite{DBLP:conf/ndss/SjostenAPS19} investigated the fingerprintability of browser extensions. Karami et al.~\cite{DBLP:conf/ndss/KaramiISP20} automated the creation and detection of browser extension fingerprinting and conducted a thorough analysis of the associated privacy threats. Jia et al.~\cite{DBLP:conf/ccs/JiaCHCSL16} examined process-based isolation in Chrome, demonstrating that existing memory vulnerabilities in Chrome's renderer can be exploited as a precursor to attacking the local system. Hausknecht et al.~\cite{DBLP:conf/dimva/HausknechtMS15} studied the implementation of Content Security Policy within Chrome extensions. Borgolte et al.~\cite{DBLP:conf/www/BorgolteF20} analyzed the impact of eight popular privacy-focused browser extensions on browser performance, finding that these extensions can enhance both user privacy and browsing experience. By analyzing the communication interfaces exposed to web applications by browser extensions, Som{\'{e}} et al.~\cite{DBLP:conf/sp/Some19} identified numerous extensions that web applications can leverage to gain access to privileged capabilities.

A substantial body of work has addressed the protection of web applications against malicious JavaScript inclusions. Arshad et al.~\cite{DBLP:conf/fc/ArshadK016} proposed a browser-based methodology for the detection of malicious third-party content inclusions and implemented it within Chromium. Xing et al.~\cite{DBLP:conf/ndss/XingCWC13} presented InteGuard to provide security against vulnerable web API integrations. ScriptInspector~\cite{DBLP:conf/sp/ZhouE15} is a specialized browser capable of intercepting, recording, and verifying third-party script accesses to critical resources against predefined security policies. Lekies et al.~\cite{DBLP:conf/uss/LekiesSWJ15} conducted a systematic investigation into the security concerns stemming from dynamic JavaScript inclusion. Phung et al.~\cite{DBLP:conf/ccs/PhungSC09} introduced a technique for transforming JavaScript code into a self-protective form based on function wrapping. Magazinius et al.~\cite{DBLP:conf/nordsec/MagaziniusPS10} expanded upon Phung's work by developing a systematic approach to mitigate the identified vulnerabilities. Meyerovich et al.~\cite{DBLP:conf/www/MeyerovichFM10} proposed object views as a user-level mechanism for fine-grained JavaScript object sharing. Google's Caja project~\cite{website:Caja} and Facebook's FBJS project~\cite{website:FBJS} adopted JavaScript function wrapping to enforce security checks.

\section{Conclusion}%
\label{sec:conclusion}

We developed and deployed a suite of tools to systematically gather JavaScript inclusions within Chrome extensions, subsequently collecting data from a corpus of {\extensionNum} Chrome extensions. Our investigation reveals the continued presence of JavaScript inclusion issues within this ecosystem. In the context of Chrome extensions, developers exhibit a tendency towards the utilization of locally sourced JavaScript resources rather than importing them from external servers. Nevertheless, a significant number of JavaScript libraries are implemented via code duplication, leading to suboptimal maintenance practices. Consequently, a considerable quantity of deprecated JavaScript libraries persists within Chrome extensions. This issue appears particularly pronounced in widely adopted Chrome extensions. Finally, we propose recommendations aimed at mitigating this identified problem.

\bibliographystyle{splncs04}
\bibliography{main}

\end{document}